\documentstyle[prl,aps,epsfig]{revtex}

\begin{document}
\large

\title{Use of non-adiabatic geometric phase for  quantum computing by nuclear magnetic resonance}
\author{Ranabir Das, S. K. Karthick Kumar and Anil Kumar$^*$ \\
        {\small \it NMR Quantum Computation and Quantum Information Group \\
Department of Physics and Sophisticated Instruments Facility, Indian Institute of Science, 
Bangalore-560012, India}}
\maketitle
\vspace{0.5cm}
\begin{abstract}
Geometric phases have stimulated researchers for its potential applications in many areas of science. One of them is
fault-tolerant quantum computation. A preliminary requisite of quantum computation is the implementation of controlled 
logic gates by controlled dynamics of qubits. In controlled dynamics, one qubit undergoes coherent
evolution and acquires appropriate phase, depending on the state of other qubits. If the evolution is geometric,
then the phase acquired depend only on the geometry of the path executed, and is robust against certain
types of errors. This phenomenon leads to an inherently fault-tolerant quantum computation.
 Here we suggest a technique of using non-adiabatic geometric phase for quantum computation, using selective excitation.
In a two-qubit system, we selectively evolve a suitable subsystem 
 where the control qubit is in state $\vert 1\rangle$, through a closed circuit. By this evolution, the target qubit 
gains a phase controlled by the state of the control qubit. Using these geometric phase gates we demonstrate implementation of 
Deutsch-Jozsa algorithm and Grover's search algorithm in a two-qubit system.
\end{abstract}

\section{Introduction}
Quantum states which differ only by a overall phase cannot be distinguished by measurements 
in quantum mechanics. Hence phases were thought to be unimportant 
until Berry made an important and interesting observation regarding the behavior of pure quantum
 systems in a slowly changing environment \cite{2}. The adiabatic theorem makes sure that, if a system is initially in an eigenstate of the
 instantaneous Hamiltonian, it remains so. When the environment (more precisely, the Hamiltonian) returns to it's initial state 
after undergoing slow changes, the system acquires a measurable phase, apart from the well known dynamical phase, 
which is purely of geometric origin \cite{2}.
Simon\cite{3} showed this to be a consequence of parallel transport in a curved space appropriate to the quantum system. 
 Berry's phase was reconsidered 
by Aharanov and Anandan, who shifted the emphasis from changes in the environment, to the motion of the pure quantum system 
itself and found that the for all the changes in the environment, the same geometric phase is obtained which is uniquely 
associated with the motion of the pure quantum system and hence enabled them to generalize Berry's phase to non-adiabatic motions
\cite{5}.
 For a  spin half particle subjected to a magnetic field $\bf B$, the non adiabatic cyclic Aharanov-Anandan
 phase is just the 
solid angle determined by the path in the projective Hilbert space \cite{5}. 

Yet another interesting discovery in the fundamentals of quantum physics was the observation that by accessing a 
large Hilbert space spanned by the linear combination of quantum states and by intelligently manipulating them, some of 
the problems intractable for classical computers can be solved efficiently \cite{rf,preskill}. 
This idea of quantum computation 
using coherent quantum mechanical systems has excited a number of research groups \cite{bou,chuangbook}. 
Various physical systems including nuclear magnetic resonance (NMR) are being examined to built a suitable physical 
device which would perform quantum information processing and quantum simulations \cite{dg,cory98,na,ernst,jo}. 
Also, the quantum correlation inherently present in the 
entangled quantum states was found to be useful for quantum computation, communication and cryptography \cite{preskill}.
Geometric quantum computing is a way of manipulating quantum states using quantum gates based on 
geometric phase shifts \cite{4,9}.
 This approach is particularly useful because of the built-in fault tolerance, which arises due to the fact that geometric phases
 depend only upon some global geometric features and it is robust against certain errors and  dephasing  \cite{4,9,10,11}.

In nuclear magnetic resonance (NMR), the acquisition of geometric phase by a spin was first verified by Pines et.al.\cite{6} in  adiabatic regime by
 subjecting a nuclear spin to an effective magnetic field which slowly sweeps a cone. 
A similar approach was adopted by Jones et.al. to demonstrate the construction of controlled phase shift gates in a
two-qubit system using adiabatic geometric phase \cite{4,9}. Pines et.al. also studied the 
geometric phase in non-adiabatic regime, namely the Aharanov-Anandan phase by NMR \cite{7}. 
They used a system of two dipolar coupled identical proton spins
 which form a three level system. A two level subsystem was made to undergo a cyclic 
evolution in the Hilbert space by applying a time-dependent magnetic field, while geometric phase was observed in the
modulation of the coherence of the other two-level subsystem \cite{7}. 
 Recently, non-adiabatic geometric phase has also been observed  for mixed states by NMR using evolution in  
tilted Hamiltonian frame \cite{mix}.
In the work reported here, we have adopted a scheme similar to that of Pines et.al. \cite{7} to demonstrate 
construction of controlled phase shift gates in a two-qubit system using non-adiabatic geometric phase by NMR. The scheme is easily
 scalable to higher qubit systems. The geometric controlled phase gates were used to implement Deutsch-Jozsa (DJ) algorithm 
\cite{deu} and Grover's search algorithm \cite{grover} in the two-qubit system. To the best of our knowledge, this is the first implementation of 
quantum algorithms using geometric phase.  

\section{non-adiabatic geometric phase gate}

Consider a two qubit system, which has four eigenstates $\vert 00\rangle$, $\vert 01\rangle$, $\vert 10\rangle$
and $\vert 11\rangle$. The two-state subsystem of $\vert 10\rangle$ and $\vert 11\rangle$ can be taken through 
a circuit enclosing a solid angle $\Omega$ \cite{7}. If the other dynamical phases are canceled during the process, these 
two states gain a non-adiabatic phase purely due to geometric topology. Since the operation is done selectively 
with the states where first qubit is in state $\vert 1 \rangle$, this acts as a controlled phase gate where the 
second qubit gains a phase only when the first qubit is $\vert 1 \rangle$ \cite{bou,chuangbook,4} . 

 The transport of the selected states through a closed circuit can be accomplished by selective excitation.
Such selective excitation can be performed by pulses having a small bandwidth which excite a selected transition in the 
spectrum and leave the others unaffected \cite{cory98,freeman,free,kd,pram1,rd1}. In the following we consider geometric phase 
acquired by two different paths in a Bloch sphere, respectively known as  slice circuit and triangular circuit \cite{7}.

\subsection{Geometric phase acquired by a slice circuit}
In a slice circuit, the state vector cuts a slice out of the Bloch sphere, Figure 1(a).
The slice circuit can be achieved by two pulses A.B=$(\pi)_{\theta}^{\vert 10\rangle \leftrightarrow \vert 11\rangle}$.
$(\pi)_{\theta+\pi+\phi}^{\vert 10\rangle \leftrightarrow \vert 11\rangle}$, where the pulses are applied from left to 
right. Here  
$(\pi)_{\theta}^{\vert 10\rangle \leftrightarrow \vert 11\rangle}$ denotes a selective $\pi$-pulse on 
$\vert 10\rangle \leftrightarrow \vert 11\rangle$ transition with phase $\theta$. The first
 $(\pi)_{\theta}^{\vert 10\rangle \leftrightarrow \vert 11\rangle}$ pulse rotates the 
polarization vector of the subsystem through $\pi$ about the an axis with azimuthal angle $\theta$ in the 
x-y plane (Fig. 1).
The vector is brought back to its original position completing a closed circuit
 by the  second $\pi$-pulse about the axis in the x-y plane with azimuthal angle $(\theta+\pi+\phi)$. 
The resulting path encloses a solid angle of 2$\phi$,
The operator of the two pulses can be calculated as,
\begin{eqnarray}
{\mathrm A.B}=&&(\pi)_{\theta}^{\vert 10\rangle \leftrightarrow \vert 11\rangle}
(\pi)_{\theta+\pi+\phi}^{\vert 10\rangle \leftrightarrow \vert 11\rangle} \nonumber \\
=&&exp[-i(I_x^{\vert 10\rangle \leftrightarrow \vert 11\rangle} cos(\theta)+
I_y^{\vert 10\rangle \leftrightarrow \vert 11\rangle} sin(\theta))\pi] \nonumber \\
&&exp[-i(I_x^{\vert 10\rangle \leftrightarrow \vert 11\rangle} cos(\theta +\pi +\phi)+
I_y^{\vert 10\rangle \leftrightarrow \vert 11\rangle} sin(\theta +\pi+\phi))\pi] \nonumber \\
=&&\pmatrix{1 & 0 & 0 & 0 \cr 0& 1& 0& 0& \cr 0& 0& 0& -sin\theta-icos\theta \cr 0& 0& sin\theta-icos\theta & 0 } \times \nonumber \\ 
&&\pmatrix{1& 0& 0& 0 \cr 0& 1& 0& 0 \cr 0& 0& 0& sin(\theta+\phi)+icos(\theta+\phi)
\cr 0& 0& -sin(\theta+\phi)+icos(\theta+\phi)& 0} \nonumber \\
&&=\pmatrix{1 & 0 & 0 & 0 \cr 0& 1& 0& 0& \cr 0& 0& e^{i\phi}& 0 \cr 0& 0& 0 & e^{-i\phi}}, 
\end{eqnarray}
 where $I_x^{\vert 10\rangle \leftrightarrow \vert 11\rangle}$ and 
$I_y^{\vert 10\rangle \leftrightarrow \vert 11\rangle}$ are the fictitious spin-1/2 operators \cite{vega} for the 
two-state subsystem  of $\vert 10\rangle$ and $\vert 11\rangle$, given by; 
\begin{eqnarray}
I_x^{\vert 10\rangle \leftrightarrow \vert 11\rangle}=\frac{1}{2}\pmatrix{0 & 0 & 0 & 0 \cr 0 & 0 & 0 & 0 \cr 
0 & 0 & 0 & 1 \cr 0 & 0 & 1 & 0} ~~~~~{\mathrm and}~~~~~
I_y^{\vert 10\rangle \leftrightarrow \vert 11\rangle}=\frac{1}{2}\pmatrix{0 & 0 & 0 & 0 \cr 0 & 0 & 0 & 0 \cr
0 & 0 & 0 & -i \cr 0 & 0 & i & 0}. 
\end{eqnarray}  

Note that the combined operator of the two pulses in Eq.[1] attributes a non-adiabatic geometric phase 
proportional to the solid angle of 
the circuit traversed. However, the phase it attributed only to the 
two states where first qubit is in state $\vert 1\rangle$. Since the selective excitation does not perturb the 
other transitions, the subsystem $\vert 00\rangle \leftrightarrow \vert 01\rangle$ do not gain any phase, 
This is analogous to the controlled phase gate where 
the second qubit acquires a phase controlled by the state of first qubit \cite{chuangbook,4}.

  To demonstrate the operation of a controlled geometric phase gate, we have taken the two qubit system of 
carbon-13 labeled chloroform ($^{13}$CHCl$_3$), where the two nuclear spins $^{13}$C and $^1$H forms the two-qubit
 system. The sample of $^{13}$CHCl$_3$ was dissolved in the solvent of CDCl$_3$, and experiments were performed 
at room temperature  at a magnetic field of B$_0$=11.2 Tesla. At this high-field the resonance frequency of proton is 
500 MHz and that of carbon is 125 MHz. The indirect spin-spin coupling (the J-coupling) 
between the two qubits is 210 Hz. Starting from equilibrium, 
the $\vert 00\rangle$ pseudopure state was prepared by spatial averaging method using the pulse sequence \cite{mix},
\begin{eqnarray}
(\pi/3)^2_x-G_z-(\pi/4)^2_x-\frac{1}{2J}-(\pi/4)^2_{-y}-G_z,
\end{eqnarray}
 where the pulses were applied on the second qubit, denoted by $2$ in superscript, which in our case is the proton 
spin. After creation of pps, a pseudo-Hadamard gate \cite{djjo,grojo} was applied on the first qubit, which in our case was $^{13}$C. The pseudo-Hadamard gate 
was implemented by a $(\pi/2)^1_y$ where the superscript denotes the qubit and the subscript denotes the phase of the pulse \cite{djjo,grojo}.
This gate creates a an uniform superposition of the first qubit $\vert 00\rangle +\vert 10\rangle$. The operation of the 
controlled phase gate would  now transform the state into $\vert 00\rangle + e^{i\phi}\vert 10\rangle$. 
For the slice circuit, the proton dynamical phase would vanish since the applied field is always orthogonal to the 
polarization vector, generating parallel transport \cite{7}. However, the carbon coherence would undergo evolution due to the 
internal Hamiltonian during the pulses. Hence the pulse sequence of the gate was incorporated into a Hahn-echo 
\cite{echo,ernstbook} sequence of the 
form $\tau-(\pi)_x-\tau$, where the pulse sequence of the gate were applied during the second $\tau$ period, as given in 
figure 1(b). The 
intermediate $(\pi)$-pulse refocuses inhomogeneity of the B$_0$ field, the chemical shift of carbon and its J-coupling to the proton. 
However, to restore the state of the first qubit altered by the $(\pi)$-pulse, 
the pulse sequence of Eq. [1] has to be supplemented by adding 
a $(\pi)^1_{-x}$ pulse (figure 1(b)), yielding the sequence:
\begin{eqnarray}
&&(\pi)^1_{x}.(\pi)_{\theta}^{\vert 00\rangle \leftrightarrow \vert 01\rangle}.
(\pi)_{\theta+\pi+\phi}^{\vert 00\rangle \leftrightarrow \vert 01\rangle}.(\pi)^1_{-x} \nonumber \\
&&=\pmatrix{0 & 0 & i & 0 \cr 0 & 0 & 0 & i \cr i & 0 & 0 & 0 \cr 0 & i & 0 & 0}.
\pmatrix{e^{i\phi} & 0 & 0 & 0 \cr 0& e^{-i\phi} & 0& 0& \cr 0& 0& 1 & 0 \cr 0& 0& 0 & 1}.
\pmatrix{0 & 0 & -i & 0 \cr 0 & 0 & 0 & -i \cr -i & 0 & 0 & 0 \cr 0 & -i & 0 & 0} \nonumber \\
&&=\pmatrix{1 & 0 & 0 & 0 \cr 0& 1& 0& 0& \cr 0& 0& e^{i\phi}& 0 \cr 0& 0& 0 & e^{-i\phi}},
\end{eqnarray}     
  where the selective pulses were applied on the $\vert 00\rangle \leftrightarrow \vert 01\rangle$ transition to 
achieve the exact form of controlled phase gate. 
The selective excitation was obtained with Gaussian shaped pulses of 13.2 ms duration. The non-adiabatic geometric phase 
was observed in the phase of $\vert 00\rangle \leftrightarrow \vert 10\rangle$ coherence. We have observed the geometric phase 
for the slice circuit with various solid angles $(2\phi)$, each time varying the phase $\phi$ of the second selective $(\pi)$-pulse. 
The corresponding spectra are given in figure 2(c), where the $\vert 00\rangle \leftrightarrow \vert 10\rangle$ shows a 
phase change of $e^{i\phi}$. For $\phi=0$, there is no phase change and the peak 
is absorptive. With increase of $\phi$, the phase of the peak changes and it  becomes dispersive for $\phi=\pi/2$, 
and subsequently, a negative absorptive for $\phi=\pi$.

  The three small lines in the spectra comes from the naturally abundant 
$^{13}$C signal of CDCl$_3$, which provide a reference. 
Since all dynamical phases due to evolution under chemical shift and J-couplings were refocused, 
the solvent $^{13}$C signal is absorptive in all the spectra. However, solute $^{13}$C signal gains phase because it is coupled to 
the protons, one of whose transition is taken through a closed circuit. 
This result thus provides a graphic display of geometric phase by non-adiabatic evolution. To accurately read the phase angle 
of each spectrum in Fig. 2(c), a zero-order phase correction was applied to the spectra in Fig. 2(c), till the observed peak became 
absorptive. The change of phase of $\vert 00\rangle \leftrightarrow \vert 10\rangle$ coherence due to geometric phase
 is plotted against the solid-angle (2$\phi$), in figure 3. The graph in figure 3 shows the high fidelity of the 
experimental implementation of the slice circuit in this case.

\subsection{Geometric phase acquired by a triangular circuit}
 In the triangular circuit, the state vector traverses a triangular path on the Bloch sphere figure 1(c) \cite{7}.
The solid angle enclosed by the triangular circuit of figure 1(c) is $\phi$.
The controlled  phase shift gate  can be implemented by the non-adiabatic phase acquired when the appropriate  
sub-system goes through this circuit.
The pulse sequence for the circuit and the corresponding operator can be calculated, 
similar to that of the sliced circuit, as
\begin{eqnarray}
A.C.B=&&(\pi/2)^{\vert 10\rangle \leftrightarrow \vert 11\rangle}_{\theta}.
(\phi)^{\vert 10\rangle \leftrightarrow \vert 11\rangle}_{z}.
(\pi/2)^{\vert 10\rangle \leftrightarrow \vert 11\rangle}_{\theta+\pi-\phi} \nonumber \\
=&&\pmatrix{1 & 0 & 0 & 0 \cr 0& 1& 0& 0& \cr 0& 0& \frac{1}{\sqrt{2}}& \frac{-sin\theta-icos\theta}{\sqrt{2}}
 \cr 0& 0& \frac{sin\theta-icos\theta}{\frac{1}{\sqrt{2}}} & \frac{1}{\sqrt{2}} } \times 
\pmatrix{1& 0& 0& 0 \cr 0& 1& 0& 0 \cr 0& 0& e^{-i\phi/2}& 0 \cr 0 &0& 0& e^{i\phi/2}} \times \nonumber \\
&&\pmatrix{1& 0& 0& 0 \cr 0& 1& 0& 0 \cr 0& 0& \frac{1}{\sqrt{2}}& \frac{sin(\theta-\phi)+icos(\theta-\phi)}{\sqrt{2}} \cr
 0& 0& \frac{-sin(\theta-\phi)+icos(\theta-\phi)}{\sqrt{2}}& \frac{1}{\sqrt{2}}} \nonumber \\
&&=\pmatrix{1 & 0 & 0 & 0 \cr 0& 1& 0& 0& \cr 0& 0& e^{-i\phi/2}& 0 \cr 0& 0& 0 & e^{i\phi/2}},
\end{eqnarray}
The intermediate $(\phi)^{\vert 10\rangle \leftrightarrow \vert 11\rangle}_{z}$ pulse can be applied by the 
composite z-pulse sequence $(\pi/2)^{\vert 10\rangle \leftrightarrow \vert 11\rangle}_{y}
(\phi)^{\vert 10\rangle \leftrightarrow \vert 11\rangle}_{-x}(\pi/2)^{\vert 10\rangle \leftrightarrow \vert 11\rangle}_{-y}$ 
\cite{lev,ranajmr}. 

In the experiments, we have chosen $\theta=3\pi/2$. The state of $\vert 00\rangle +\vert 10\rangle$ was prepared and 
 then the pulse sequence of figure 1(d) was applied. Similar to the slice circuit, the sequence was incorporated 
in a Hahn-echo and the pulses were applied on the $\vert 00\rangle \leftrightarrow \vert 01\rangle$ transition.
The operator of Eq.[5] transforms $\vert 00\rangle +\vert 10\rangle$ to $\vert 00\rangle +e^{-i\phi/2}\vert 10\rangle$.  
The phase of the $\vert 00\rangle \leftrightarrow \vert 10\rangle$ was observed for various $\phi$, by changing the angle 
of the z-pulse and the phase of the last pulse in Eq.[5]. The spectra are given in figure 4. Once again, the peak changes 
from absorptive to dispersive and then to a negative absorptive in correspondence with the change of $\phi$. 
  
   However, there are two major 
differences between the spectra of figure 2(c) and 4(c). Note that after the phase gate, the state of the system is 
$\vert 00\rangle +e^{i\phi}\vert 10\rangle$ for slice circuit and $\vert 00\rangle +e^{-i\phi/2}\vert 10\rangle$ 
for triangular circuit. This is because the solid angle of the slice circuit if 2$\phi$, whereas that of the 
triangular circuit is $\phi$. Hence,  in the slice circuit the coherences become a negative 
absorptive for $\phi=\pi$, 
whereas  in the triangle circuit the same observation is obtained for $\phi=2\pi$. Moreover, the phase of the pulses 
corresponding to the triangle circuit is chosen such that the sign of phase is opposite to that of the 
slice circuit. This difference is clearly reflected in the sign of the coherences between figure 2(c) and 4(c).  
A plot of the absolute value of observed phase change 
against solid angle is given in figure 5, whose high fidelity validate the use of such gates for quantum computing.

\section{Deutsch-Jozsa algorithm}
 Deutsch-Jozsa (DJ) algorithm provides a demonstration of the advantage of quantum superpositions over 
classical computing \cite{deu}.
The DJ algorithm determines the type of an unknown function when it is either constant or balanced. In the simplest
case, $f(x)$ maps a single bit to a single bit. The function is called constant if $f(x)$ is
independent of $x$ and it is balanced if  $f(x)$ is zero for one value of $x$ and unity for the other
value. For N qubit system,  $f(x_1,x_2,...x_N)$ is constant if it is independent of $x_i$ and balanced if it is zero
for half the values of $x_i$ and unity for the other half. Classically it requires ($2^{N-1}+1$)
function calls to check if $f(x_1,x_2,...x_N)$ is constant or balanced.
 However the DJ algorithm would require only a single function call \cite{deu}. The Cleve version of DJ algorithm
 implemented by using a unitary transformation  by the propagator
$U_f$ while adding an extra qubit, is given by \cite{cleve},
\begin{eqnarray}
\vert x_1,x_2,...x_N\rangle \vert x_{N+1}\rangle \stackrel{U_f}{\longrightarrow}
\vert x_1,x_2,...x_N\rangle \vert x_{N+1}\oplus f(x_1,x_2,...x_N)\rangle
\end {eqnarray}
The four possible functions for the single-bit DJ algorithm are $f_{00}$, $f_{11}$, 
$f_{10}$ and $f_{01}$. $f_{00}(x)=0$ for $x=$0 or 1, $f_{11}(x)=1$ for $x=$0 or 1, 
$f_{10}(x)=$1 or 0 corresponding to  $x=$0 or 1, while $f_{01}(x)=$0 or 1 corresponding to $x=$0 or 1.
The unitary transformations corresponding to the four possible propagators $U_f$ are
\begin{eqnarray}
U_{f_{00}}=\pmatrix{1&0&0&0\cr 0&1&0&0\cr 0&0&1&0\cr 0&0&0&1},~~~~~~~~~~
U_{f_{11}}=\pmatrix{0&1&0&0\cr 1&0&0&0\cr 0&0&0&1\cr 0&0&1&0}, \nonumber \\
\nonumber \\
U_{f_{10}}=\pmatrix{1&0&0&0\cr 0&1&0&0\cr 0&0&0&1\cr 0&0&1&0},~~~~~~~~~~
U_{f_{01}}=\pmatrix{0&1&0&0\cr 1&0&0&0\cr 0&0&1&0\cr 0&0&0&1}.
\end{eqnarray}
For higher qubits the functions are easy to evaluate using Eq.[6].
DJ-algorithm has been demonstrated using dynamic phase by several research groups \cite{djjo,djchu,ka1,pram,ranajcp}.

 The quantum circuit for single-bit Cleve version of DJ algorithm is given in figure 6(a) \cite{djchu}.
The algorithm starts with $\vert 00\rangle$ pseudopure state. The pair of pseudo-Hadamard gates 
$(\pi/2)^1_y(\pi/2)^2_{-y}$ create superposition of the form
$[(\vert 0\rangle + \vert 1\rangle)/\sqrt{2}][(\vert 0\rangle - \vert 1\rangle)/\sqrt{2}]$.
Then the operator $U_f$ is applied. When the function is constant, i.e. $f(0)=f(1)$, the input qubit is in the state  
$(\vert 0\rangle + \vert 1\rangle)/\sqrt{2}$, else 
the function is balanced in which case it is in the state $(\vert 0\rangle - \vert 1\rangle)/\sqrt{2}$. 
Thus, the answer is stored in the
relative phase between the two states of the input qubit. A final pair of pseudo-Hadamard gates
$(\pi/2)^1_{-y}(\pi/2)^2_{y}$ converts the superposition back into the eigenstates.
The work qubit comes back to state $\vert 0\rangle$, where as the input qubit becomes $\vert 0\rangle$ or $\vert 1\rangle$
corresponding to the function being constant or balanced.

The operator of $U_{f_{00}}$ is identity matrix and corresponds to no operation. The operator of $U_{f_{11}}$
can be achieved by a $(\pi)_x$ pulse on the second qubit. In this experiment, unlike section II, we label proton as 
the first qubit and carbon as the second qubit, and consequently the $(\pi)_x$ pulse was applied on the carbon. 
The $U_{f_{10}}$ operator is a controlled-NOT gate which flips the second qubit when the first qubit is $\vert 1\rangle$.
This gate can be achieved by a controlled phase gate sandwiched between two pseudo-Hadamard gates on the second qubit [],
$U_{f_{10}}=h-C_{11}(\pi)-h^{-1}$, where the controlled phase gate is of the form, 
\begin{eqnarray}
C_{11}(\phi)=\pmatrix{1& 0& 0& 0 \cr 0& 1& 0& 0 \cr 0& 0& 1& 0 \cr 0& 0& 0& e^{i\phi}}. 
\end{eqnarray} 
 This precise form of controlled phase gate can be achieved by a recursive use of the phase gates demonstrated 
in section II. Since the gate A.B given in Eq.[1] attributes a phase $e^{i\phi}$ to the state $\vert 10\rangle$ and $e^{-i\phi}$
to the state $\vert 11\rangle$, we denote this gate as $C_{10}(\phi).C_{11}(-\phi)$, where 
\begin{eqnarray}
A.B=[C_{10}(\phi).C_{11}(-\phi)]=\pmatrix{1& 0& 0& 0 \cr 0& 1& 0& 0 \cr 0& 0& e^{i\phi}& 0 \cr 0& 0& 0& e^{-i\phi}}.
\end{eqnarray}
 The phase gate $C_{11}(\phi)$ can be constructed by a suitable combination of these gates,
\begin{eqnarray}
&&[C_{00}(-\phi/4).C_{10}(\phi/4)] \times 
[C_{01}(-\phi/4).C_{11}(\phi/4)] \times [C_{10}(-\phi/2).C_{11}(\phi/2)] \nonumber \\
&&= \pmatrix{e^{-i\phi/4}& 0& 0& 0 \cr 0& 1& 0& 0 \cr 0& 0& e^{i\phi/4}& 0 \cr 0& 0& 0& 1} \times
\pmatrix{1& 0& 0& 0 \cr 0& e^{-i\phi/4}& 0& 0 \cr 0& 0& 1& 0 \cr 0& 0& 0& e^{i\phi/4}} \times
\pmatrix{1& 0& 0& 0 \cr 0& 1& 0& 0 \cr 0& 0& e^{-i\phi/2}& 0 \cr 0& 0& 0& e^{i\phi/2}} \nonumber \\
&&=\pmatrix{e^{-i\phi/4}& 0& 0& 0 \cr 0& e^{-i\phi/4}& 0& 0 \cr 0& 0& e^{-i\phi/4}& 0 \cr 0& 0& 0& e^{i3\phi/4}}
=e^{-i\phi/4}\pmatrix{1& 0& 0& 0 \cr 0& 1& 0& 0 \cr 0& 0& 1& 0 \cr 0& 0& 0& e^{i\phi}}=e^{-i\phi/4}C_{11}(\phi).
\end{eqnarray}

 Note that if performed in fault-tolerant manner by  using non-adiabatic geometric phase, 
the first gate requires a rotation of the transition 
$\vert 00\rangle \leftrightarrow \vert 10\rangle$ through a closed circuit. We have used the slice circuit, where it requires 
a sequence of two $\pi$-pulses, $(\pi)_{\theta}^{\vert 00\rangle \leftrightarrow \vert 10\rangle}
(\pi)_{\theta+\pi-\phi/4}^{\vert 00\rangle \leftrightarrow \vert 10\rangle}$. Similarly, the second phase gate of Eq.[7] can be 
achieved by the pulse sequence $(\pi)_{\theta}^{\vert 01\rangle \leftrightarrow \vert 11\rangle}
(\pi)_{\theta+\pi-\phi/4}^{\vert 01\rangle \leftrightarrow \vert 11\rangle}$. Note that these two sequence is require 
pulsing of both the transitions of first qubit, $\vert 00\rangle \leftrightarrow \vert 10\rangle$ for the first gate and 
$\vert 01\rangle \leftrightarrow \vert 11\rangle$ for the second. 
 Hence, they can be performed simultaneously by a couple spin-selective pulses $(\pi)_{\theta}^1(\pi)_{\theta+\pi-\phi/4}^1$, 
where the pulses are applied on the first qubit (denoted by superscript). Thus, 
\begin{eqnarray}
C_{11}(\phi)=(\pi)_{\theta}^1.(\pi)_{\theta+\pi-\phi/4}^1.(\pi)_{\theta}^{\vert 10\rangle \leftrightarrow \vert 11\rangle}.
(\pi)_{\theta+\pi-\phi/2}^{\vert 10\rangle \leftrightarrow \vert 11\rangle}.
\end{eqnarray}
 In this case $\phi=\pi$, and we have chosen $\theta=3\pi/2$. 
 The last two pulses are however transition selective pulses, which were incorporated into a refocusing sequence, 
$\tau-(\pi/2)^1_x-\tau-(\pi/2)^2_x-\tau-(\pi/2)^1_x-\tau-(\pi/2)^2_x$, where the selective pulses were applied in the 
last $\tau$ period, and the pulses were applied on the $\vert 00\rangle \leftrightarrow \vert 01\rangle$ transition.
It may be noted that the triangular circuit 
could have also used for the same purpose. The pseudo-Hadamard pulses on second qubit were achieved by 
$h=(\pi/2)^2_y$ and $h^{-1}=(\pi/2)^2_{-y}$ pulses. 

  The operator of  $U_{f_{01}}$ can be implemented in the similar manner by $h-C_{00}(\pi)-h^{-1}$, where 
$C_{00}(\phi)$ can be implemented by 
\begin{eqnarray}
C_{00}(\phi)=(\pi)_{\theta}^1.(\pi)_{\theta+\pi+\phi/4}^1.(\pi)_{\theta}^{\vert 10\rangle \leftrightarrow \vert 11\rangle}
.(\pi)_{\theta+\pi+\phi/2}^{\vert 00\rangle \leftrightarrow \vert 01\rangle}.
\end{eqnarray}  
 The equilibrium spectrum of the two qubits are given in figure 6(b).
 After creating the superposition from pps, 
applying the various $U_f$, and applying the last set of $(\pi/2)$ pulses, 
the spectra of proton and carbon were recorded in two different experiments by selective $(\pi/2)$ pulses after a gradient. The 
spectra corresponding to various functions are given in figure 6(c), (e), (g) and (i). 
The intensities of the peaks in the spectra provide a measure of the diagonal elements of the density matrix.
The complete tomographed \cite{chutomo,rd} density matrices in each case is given in figures 6(d), (f), (h) and (j). 
When $U_{f_{00}}$ and $U_{f_{11}}$ are implemented, the 
final state is $\vert 00\rangle$, and since the state of input qubit is $\vert 0\rangle$, the corresponding functions 
$f_{00}$ and $f_{11}$ are inferred to be constant. Whereas in the case of $U_{f_{01}}$ and $U_{f_{10}}$, the final state of the system in 
$\vert 10\rangle$. The state of input qubit being $\vert 1\rangle$, the corresponding functions $f_{01}$ and $f_{10}$ are 
balanced. Theoretically, it is expected that the density matrices will have only the populations 
corresponding to the final pure states. There were however errors due to r.f. inhomogeneity  and relaxation.
 The deviation from the expected results are within 13$\%$.

\section{Grover's search algorithm}
 Grover's search algorithm can search an unsorted database of size N in $O(\sqrt{N})$ steps while a classical search would require
$O(N)$ steps \cite{grover}. Grover's search algorithm has been earlier demonstrated by several 
workers by NMR, all using dynamic phase \cite{grojo,grochu,ap,ranacpl,ranajcp}.
The quantum circuit for implementing Grover's search algorithm on two qubit system is given in figure 7(a).
The algorithm starts from a $\vert 00\rangle$ pseudopure state.  A uniform superposition of all states are created by
the initial Hadamard gates $(H)$. Then the sign of the searched state $``x"$ is inverted by the oracle through the operator
\begin{eqnarray}
U_x=I-2\vert x\rangle \langle x \vert,
\end{eqnarray} 
where $U_x$ is a controlled phase shift gate $C_{x}(\pi)$. $C_{11}(\pi)$ and $C_{00}(\pi)$ gates were implemented by the 
pulse sequences given in Eq.[11] and [12] respectively. The oracle for the other two states $\vert 01\rangle$ and $\vert 10\rangle$ 
were implemented by the sequences,
\begin{eqnarray}
C_{01}(\phi)=(\pi)_{\theta}^1.(\pi)_{\theta+\pi+\phi/4}^1.(\pi)_{\theta}^{\vert 10\rangle \leftrightarrow \vert 11\rangle}
.(\pi)_{\theta+\pi-\phi/2}^{\vert 00\rangle \leftrightarrow \vert 01\rangle}, \nonumber \\
C_{10}(\phi)=(\pi)_{\theta}^1.(\pi)_{\theta+\pi-\phi/4}^1.(\pi)_{\theta}^{\vert 10\rangle \leftrightarrow \vert 11\rangle}.
(\pi)_{\theta+\pi+\phi/2}^{\vert 10\rangle \leftrightarrow \vert 11\rangle},
\end{eqnarray}
 where $\phi=\pi$, as required in our case. 

An inversion about mean is performed on all the states by a diffusion operator $HU_{00}H$ \cite{grover}, where
\begin{eqnarray}
U_{00}=I-2\vert 00\rangle \langle 00 \vert,
\end{eqnarray}
 where $U_{00}$ is nothing but $C_{00}(\pi)$, and was implemented by the pulse sequence of Eq.[12].
 For an N-sized database the algorithm requires $O(\sqrt{N})$ iterations of $U_x HU_{00}H$ \cite{grover}. For a 2-qubit system 
with four states, only one iteration is required \cite{grojo,grochu}.
We have created a $\vert 00\rangle$ pseudopure state using Eq.[3] and applied the quantum circuit of figure 7(a), for 
$\vert x \rangle=\vert 00 \rangle$, $\vert 01 \rangle$, $\vert 10 \rangle$ and $\vert 11 \rangle$. Finally, 
the spectra of proton and carbon were recorded individually in two different experiments by selective $(\pi/2)$ 
pulses after a gradient.
The complete tomographed density matrices in each case is given in figures 7(d), (f), (h) and (j).
In each case, the searched state $\vert x\rangle$ was found to be with highest probability. 
Ideally in a two-qubit system, probability should exist only in the searched state, and there should be no coherences. 
Experimentally however, other states were also found with low probability, and some coherences were found in the  
off-diagonal elements of the density matrix. These errors are mainly due to relaxation and imperfection of pulses caused by r.f. inhomogeneity.
Imperfection of r.f. pulses can cause imperfect refocusing of dynamic phase.
However, it was found that setting the duration of selective pulses to multiples of 
(2/J) yielded better results. We have used 13.2ms (6/J) duration Gaussian shaped pulses. The maximum errors in the diagonal elements 
are within 10$\%$ and that in the off-diagonal elemenst are within 15$\%$. 

\section{conclusion}
 A technique of using non-adiabatic geometric phase for quantum computing by NMR is demonstrated. 
The technique uses selective excitation of subsystems, and is easily scalable to higher qubit systems provided 
the spectrum is well resolved. Since the non-adiabatic geometric phase does not depend on the details of the 
path traversed, it is insusceptible to certain errors yielding inherently fault-tolerant quantum computation \cite{fcomp1,fcomp2}.   
The controlled geometric phase gates were also used to implement DJ-algorithm and 
Grover's search algorithm in a two-qubit system. Implementation of fault-tolerant 
controlled phase gates using adiabatic geometric phase demands that the evolution should be 'adiabatic', which requires 
long experimental time. To avoid decoherence, use of non-adiabatic geometric phase might be utile.  
\section{acknowledgment}
The authors thank K.V. Ramanathan for useful discussions. The use of DRX-500 NMR spectrometer funded by the Department of
Science and Technology (DST), New Delhi, at the Sophisticated Instruments Facility,
Indian Institute of Science, Bangalore, is gratefully acknowledged.
AK acknowledges ``DAE-BRNS" for senior scientist support and DST for a research grant for
"Quantum Computing by NMR".

$^*$DAE/BRNS Senior Scientist, e-mail: anilnmr@physics.iisc.ernet.in

\pagebreak
{\Large Figure captions}\\
 Figure 1: (a) The transport of a selected subsystem of two states $\vert r\rangle$ and $\vert s\rangle$ through 
slice circuit \cite{7}. (b) Corresponding pulse sequence. A and B are transition selective pulses incorporated into a Hahn-echo \cite{7}, where 
A= $(\pi)_{\theta}^{\vert r\rangle \leftrightarrow \vert s\rangle}$ and B=
$(\pi)_{\theta+\pi+\phi}^{\vert r\rangle \leftrightarrow \vert s\rangle}$. The path of the polarization vector 
under applied r.f. pulses is shown as A and B in (a). Due to the A= $(\pi)_{\theta}^{\vert r\rangle \leftrightarrow \vert s\rangle}$
 pulse, the polarization traverses a path from +z to -z. It comes back to +z along a different path if it is rotated 
about an axis with azimuthal angle $(\theta+\pi+\phi)$, therby enclosing a sliced circuit of solid angle $2\phi$. 
 (c) The transport of a selected subsystem of two states $\vert r\rangle$ and $\vert s\rangle$ through
a triangular circuit \cite{7}. (d) Pulse sequence for implementation of the triangular circuit given in (c). 
A=$(\pi/2)^{\vert r\rangle \leftrightarrow \vert s\rangle}_{\theta}$, C=
$(\phi)^{\vert r\rangle \leftrightarrow \vert s\rangle}_{z}$ and B=
$(\pi/2)^{\vert r\rangle \leftrightarrow \vert s\rangle}_{\theta+\pi-\phi}$.
The polarization vector is flipped to xy-plane by $(\pi/2)^{\vert r\rangle \leftrightarrow \vert s\rangle}_{\theta}$, rotated about 
z-axis by $(\phi)^{\vert r\rangle \leftrightarrow \vert s\rangle}_{z}$ and brought back to the z-axis by a $(\pi/2)^
{\vert r\rangle \leftrightarrow \vert s\rangle}$ rotation about $(\theta+\pi-\phi)$. The solid angle enclosed by the circuit is $\phi$. \\ 

 Figure 2: Observation of non-adiabatic geometric phase when the subsystem is taken through the slice circuit of 
figure 1(a). (a)  Equilibrium $^{13}$C spectrum of $^{13}$CHCl$_3$. The three 
small lines are from the natural abundant $^{13}$C signal from the solvent of CDCl$_3$.
(b) The $^{13}$C coherence of the prepared $\vert 00\rangle +\vert 10\rangle$ state. 
(c) The $^{13}$C coherence of the state $\vert 00\rangle +e^{i\phi}\vert 10\rangle$. Since the subsystem 
goes through a closed circuit, the coherence gains a phase of purely geometric origin, the magnitude of which is  proportional to the 
solid angle $(2\phi)$ of the circuit. It may be noted that since all dynamical phases are refocused, the solvent signal from CDCl$_3$
is always absorptive irrespective of $\phi$. The solute ($^{13}$CHCl$_3$) signal on the other hand gains a phase of $\phi$. 
The resulting signal changes shape from pure absorptive for $\phi=0$, to intermediate phase for arbitrary $\phi$,
 dispersive for $\phi=\pi/2$ and to absorptive (negative sign) for $\phi=\pi$. \\

Figure 3: The geometric phase gained by the $^{13}$C coherence in Fig 2(c) is plotted against the corresponding solid angle $(2\phi)$.
The observed phase closely matches the expected.\\

Figure 4: Observation of non-adiabatic geometric phase when the subsystem is taken through the triangular circuit of
figure 1(c). (a)  Equilibrium $^{13}$C spectrum of $^{13}$CHCl$_3$.
(b) The $^{13}$C coherence of the prepared $\vert 00\rangle +\vert 10\rangle$ state.
 (c) The $^{13}$C coherence of the state $\vert 00\rangle +e^{-i\phi/2}\vert 10\rangle$, after the 
subsystem goes through the triangular closed circuit. 
The coherence changes
from absorptive to dispersive and then to a negative absorptive with the change of $\phi$.
It may be noted that sign of phase of the coherence is opposite to that of 2(c).\\

Figure 5: A plot of the absolute value of observed geometric phase gained by the $^{13}$C coherence in Fig 4(c) 
is plotted against the corresponding solid angle $(\phi)$. The plot demonstrates high fidelity of such gates. \\

 Figure 6: Implementation of DJ-algorithm using non-adiabatic geometric phase in the two-qubit system of 
$^{13}$CHCl$_3$. (a) Quantum circuit for implementation of DJ-algorithm in a 
two-qubit system \cite{djchu}. h=$(\pi/2)_y$ and h$^{-1}$=$(\pi/2)_{-y}$ 
(b) Equilibrium $^1$H and $^{13}$C spectra. (c) The $^1$H and $^{13}$C spectra obtained after completion 
of the quantum circuit of (a) for $U_{f_{00}}$, and application of a gradient pulse followed by
 $(\pi/2)$ pulses on $^1$H and $^{13}$C individually. (d) The complete tomographed density matrix \cite{chutomo}. The 
real and imaginary parts of the density matrix are given separately with the imaginary part being magnified 
five times ($\times 5$).     
(e), (g) and (i) are respective spectra obtained after $U_{f_{11}}$, $U_{f_{01}}$ and $U_{f_{10}}$. 
(f), (h) and (j) are the corresponding tomographed density matrices. For constant cases (c) and (e), 
the final state is $\vert 00\rangle$, as shown in (d) and (f). For balanced cases (g) and (i), 
the final state in $\vert 10\rangle$, as shown in (h) and (i). The diagonal elements have a 
fidelity of $95\%$, while off-diagonal parts have a fidelity of $87\%$. \\

 Figure 7: Implementation of Grover's search algorithm using non-adiabatic geometric phase in the two-qubit system of
$^{13}$CHCl$_3$. (a) Quantum circuit for implementation of Grover's search algorithm in a
two-qubit system \cite{grochu,grojo}. The $U_x$ and $U_{00}$ phase gates were implemented by non-adiabatic geometric phase 
using selective excitation by 13.2 ms $(6/J)$ long Gaussian shaped pulses.
(b) Equilibrium $^1$H and $^{13}$C spectra. (c) The $^1$H and $^{13}$C spectra obtained after completion
of the quantum circuit of (a) for $\vert x\rangle=\vert 00\rangle$, and application of a gradient pulse followed by
 $(\pi/2)$ pulses on $^1$H and $^{13}$C individually. The intensities of the various lines in the spectrum gives the populations 
of the density matrix. (d) The complete tomographed density matrix \cite{chutomo} after implementation of the quantum circuit (a) 
for $x=\vert00\rangle$.
The real and imaginary parts of the density matrix are given separately and the imaginary part is magnified
five times ($\times 5$).
(e), (g) and (i) are the spectra obtained when $\vert x\rangle=\vert 01\rangle$, $\vert x\rangle=\vert 10\rangle$ 
and $\vert x\rangle=\vert 11\rangle$. In each case the searched state $\vert x\rangle$ was found with highest probability 
after implementation of the search algorithm with a fidelity more than 85$\%$.

\pagebreak
\begin{figure}
\epsfig{file=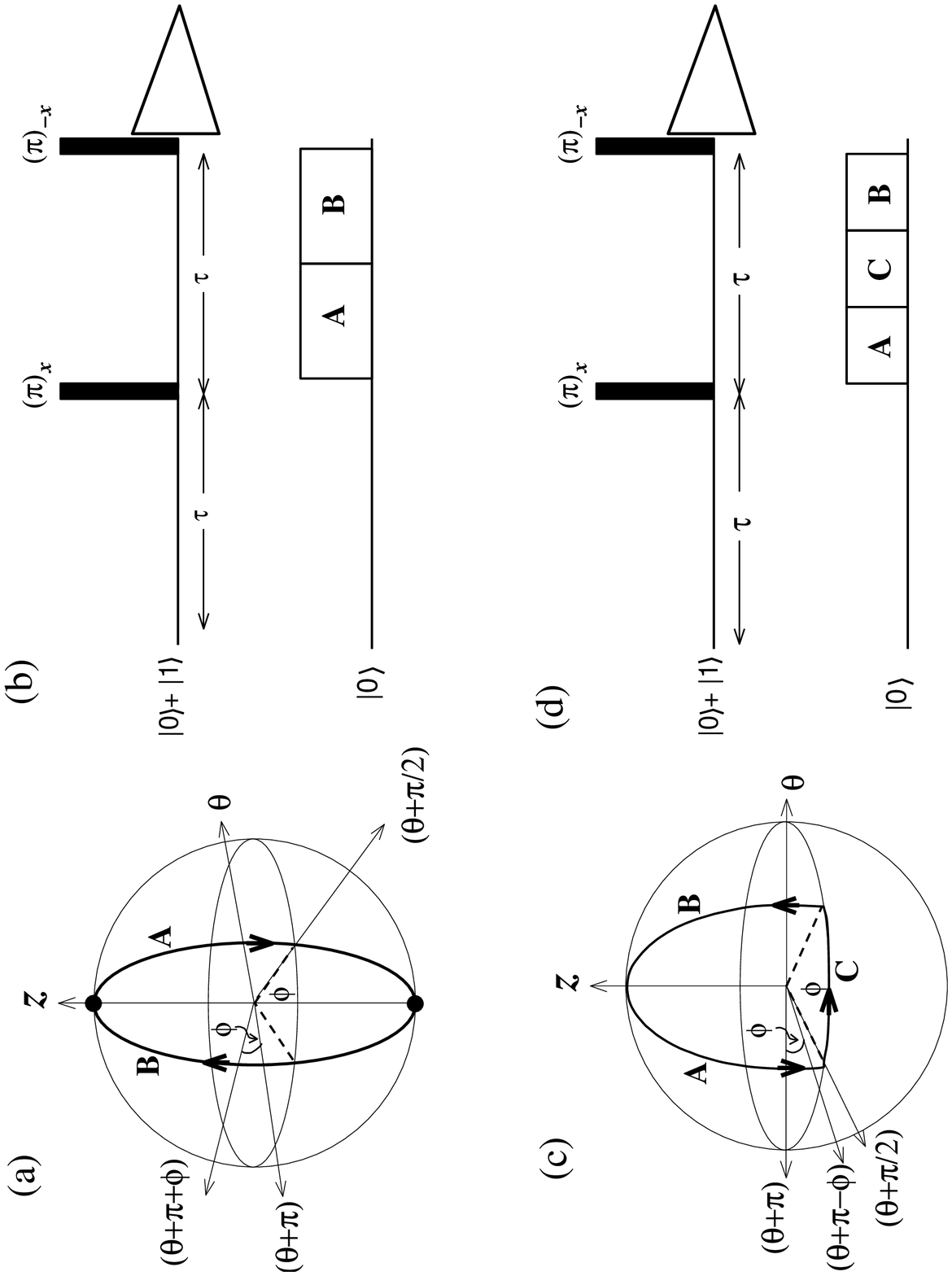,height=21cm}
\end{figure}
\hspace{5cm}
{\Large Figure 1}

\pagebreak
\begin{center}
\begin{figure}
\epsfig{file=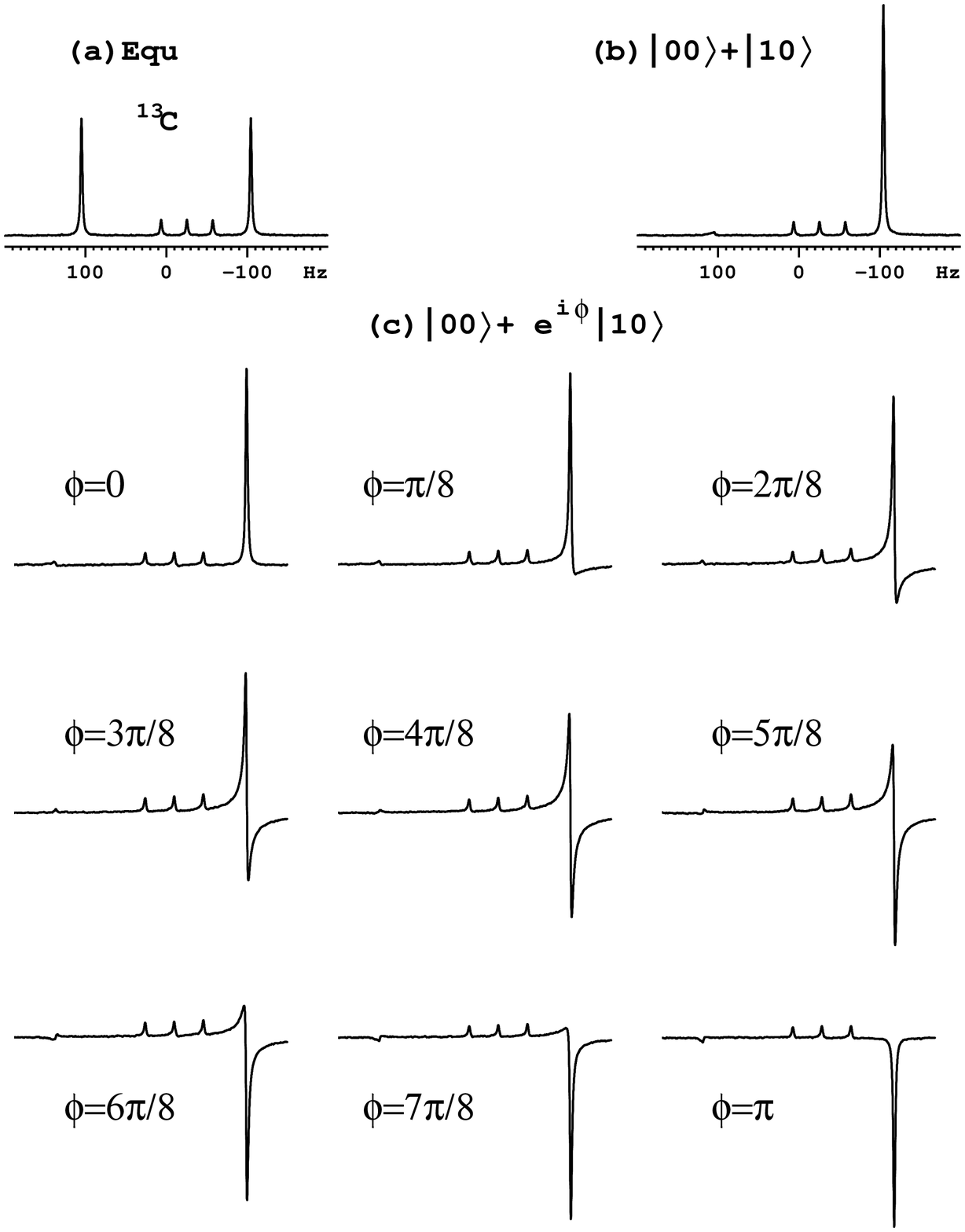,height=21cm}
\end{figure}
{\huge Figure 2}
\end{center}
\pagebreak
\begin{center}
\begin{figure}
\epsfig{file=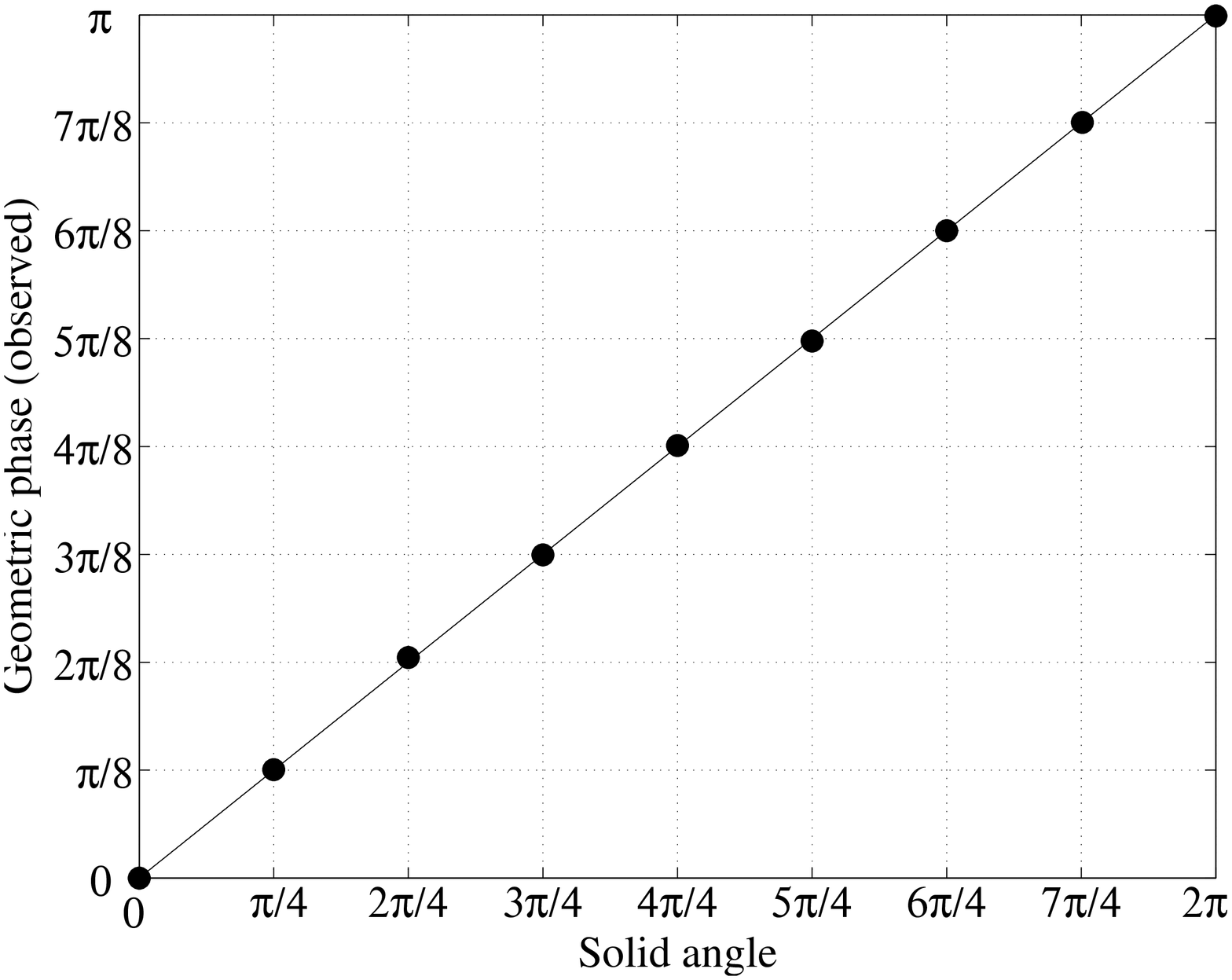,height=14cm,angle=270}
\end{figure}
{\huge Figure 3}
\end{center}

\pagebreak
\begin{figure}
\epsfig{file=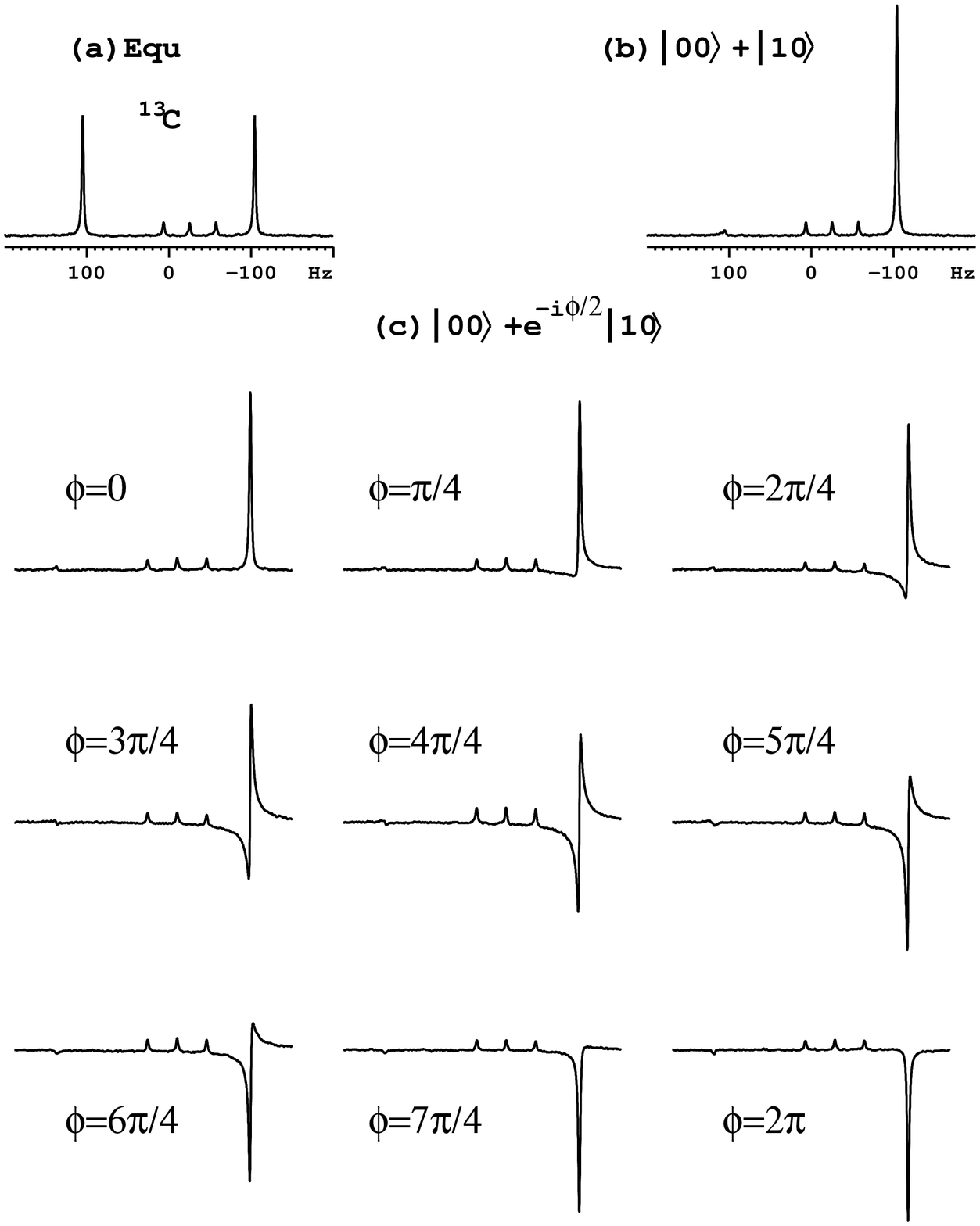,height=21cm}
\end{figure}
\hspace{5cm}
{\huge Figure 4}
\pagebreak
\begin{center}
\begin{figure}
\epsfig{file=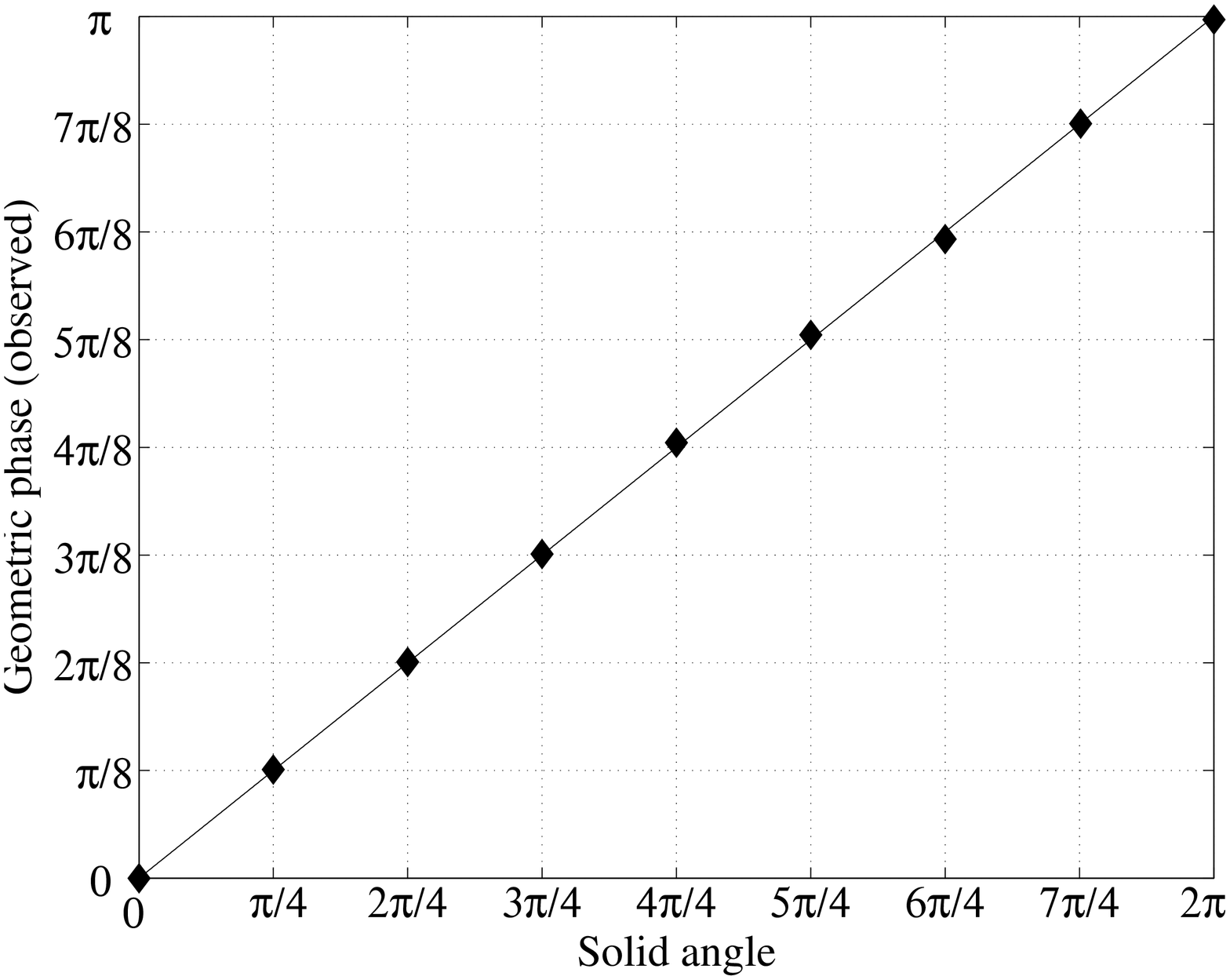,height=14cm,angle=270}
\end{figure}
{\huge Figure 5}
\end{center}

\pagebreak

\begin{figure}
\vspace*{-3cm}
\epsfig{file=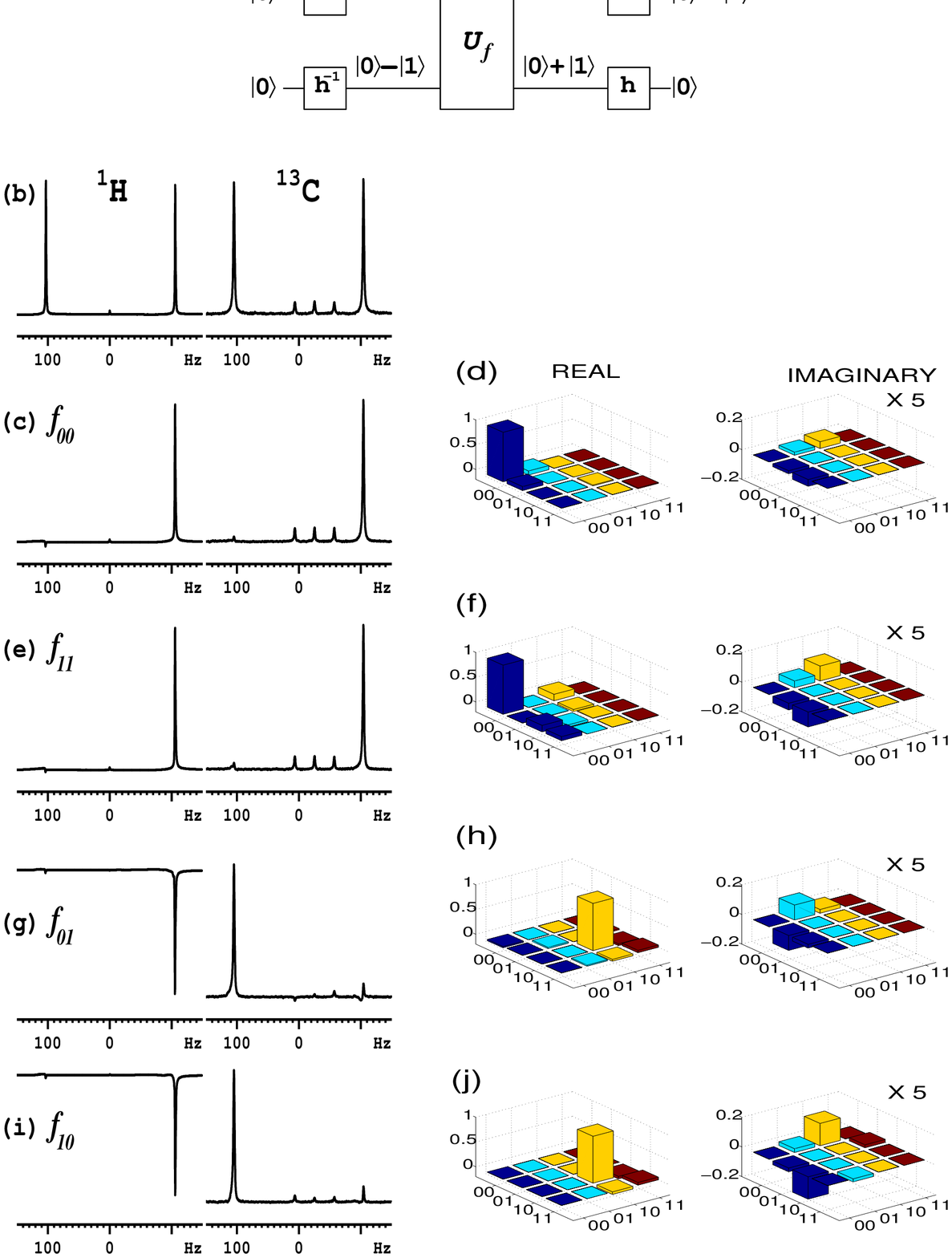,height=24cm}
\end{figure}
\hspace{7cm}
{\huge Figure 6}
\pagebreak

\begin{figure}
\vspace*{-3cm}
\epsfig{file=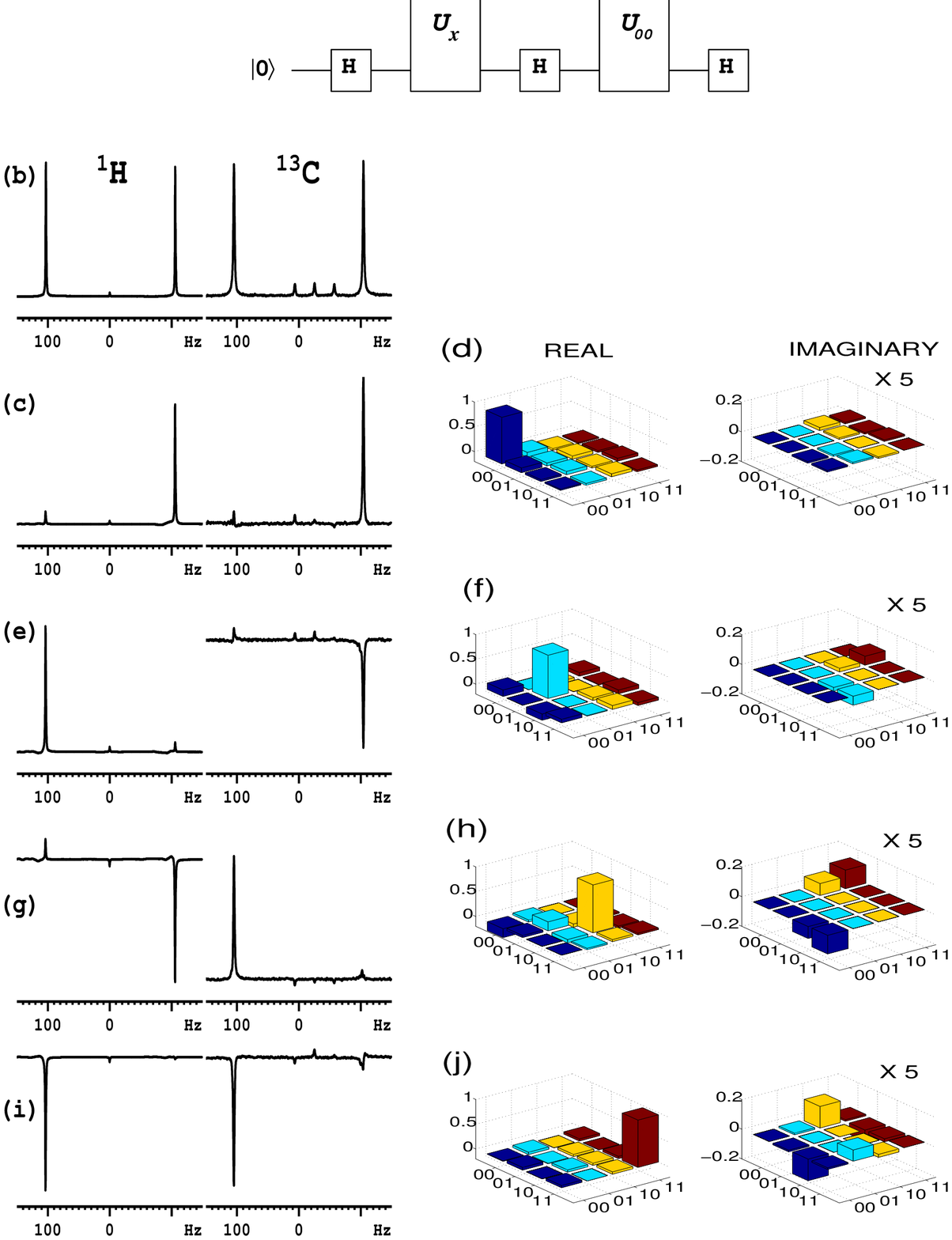,height=24cm}
\end{figure}
\hspace{7cm}
{\huge Figure 7}
\end{document}